\documentclass[conference]{IEEEtran}
\IEEEoverridecommandlockouts
\usepackage{cite}
\usepackage{amsmath,amssymb,amsfonts}
\usepackage{algorithmic}
\usepackage{graphicx}
\usepackage{textcomp}
\usepackage{xcolor}
\usepackage{xspace}
\usepackage{url}
\usepackage{subcaption}
\usepackage{mwe}
\usepackage{colortbl}
\usepackage{hyperref}

\newboolean{showcomments}
\setboolean{showcomments}{true}  % change to false to hide comments and TODOs
\ifthenelse{\boolean{showcomments}}
  {\newcommand{\nb}[2]{
    \fbox{\bfseries\sffamily\scriptsize#1}
    {\sf\small$\blacktriangleright$\textit{#2}$\blacktriangleleft$}
   }
  }
  {\newcommand{\nb}[2]{}
  }

\usepackage[shortlabels]{enumitem}  % for \begin{enumerate}[(a)]

\definecolor{added2}{HTML}{000000}
\definecolor{removed}{HTML}{DC3220}
\definecolor{added}{HTML}{000000}

\newcommand{\ie}{\emph{i.e.,}\xspace}
\newcommand{\eg}{\emph{e.g.,}\xspace}

\newcommand{\etal}{\emph{et~al.}\xspace}

\newcommand{\figref}[1]{Figure~\ref{#1}\xspace}

\newcommand{\tabref}[1]{Table~\ref{#1}\xspace}

\newcommand{\ourTool}{\tool{Currante}}
\newcommand{\tool}[1]{{\sc #1}\xspace}

\newcommand{\RQ}[1]{RQ$_{\textbf{#1}}$\xspace}
\newcommand\rev[2]{{#2}\xspace}

\definecolor{boxdarkgray}{RGB}{66,66,66}
\definecolor{boxlightgray}{RGB}{238,238,238}

\def\BibTeX{{\rm B\kern-.05em{\sc i\kern-.025em b}\kern-.08em
    T\kern-.1667em\lower.7ex\hbox{E}\kern-.125emX}}
\begin{document}

\title{Understanding Specification-Driven Code Generation with LLMs: An Empirical Study Design\textsuperscript{\small *}\thanks{\textsuperscript{*}This paper is a Stage 1 Registered Report. The study protocol and analysis plan were peer reviewed and accepted at SANER 2026 with a Continuity Acceptance (CA) score for Stage 2.}}

\author{
    \IEEEauthorblockN{
        Giovanni Rosa$^\dagger$,
        David Moreno-Lumbreras$^\dagger$,
        Gregorio Robles$^\dagger$,
        Jesús M. González-Barahona$^\dagger$
    }

    \vspace{0.5em}

    \IEEEauthorblockA{
        $^\dagger$Escuela de Ingeniería de Fuenlabrada, Universidad Rey Juan Carlos, Fuenlabrada, Spain
    }
}

\maketitle

\begin{abstract}
Large Language Models (LLMs) are increasingly integrated into software development workflows, yet their behavior in structured, specification-driven processes remains poorly understood. This paper presents an empirical study design using \textit{CURRANTE}, a Visual Studio Code extension that enables a human-in-the-loop workflow for LLM-assisted code generation. The tool guides developers through three sequential stages—Specification, Tests, and Function—allowing them to define requirements, generate and refine test suites, and produce functions that satisfy those tests. Participants will solve medium-difficulty problems from the \emph{LiveCodeBench} dataset, while the tool records fine-grained interaction logs, effectiveness metrics (e.g., pass rate, all-pass completion), efficiency indicators (e.g., time-to-pass), and iteration behaviors. The study aims to analyze how human intervention in specification and test refinement influences the quality and dynamics of LLM-generated code. The results will provide empirical insights into the design of next-generation development environments that align human reasoning with model-driven code generation.
\end{abstract}


%%%%%%%%%%%%%%%%%%%%%%%%%%%%%%%%%%%%%%%%
%%%%%%%%%%%%%%%%%%%%%%%%%%%%%%%%%%%%%%%%
\section{Introduction}
\label{sec:intro}
%%%%%%%%%%%%%%%%%%%%%%%%%%%%%%%%%%%%%%%%
%%%%%%%%%%%%%%%%%%%%%%%%%%%%%%%%%%%%%%%%

% \color{blue}
\rev{R1.1}{}
Test-Driven Development (TDD) has long been a fundamental paradigm in software development, but the introduction of Large Language Models (LLMs) is paving the way for new approaches to coding assistance, providing a potential transformation in how developers write and verify code.

The current most-used assistance for code generation is provided via code completion plugins, such as GitHub Copilot~\cite{copilot2022productivity}, or fully-featured AI-powered IDEs, such as Cursor.\footnote{\url{https://cursor.com/}}.
These tools offer powerful automation of the coding process leaveraging natural language instructions, shifting from the traditional code-centric development approach to a more abstract specification-driven process, namely \emph{Spec-Driven Development} (SDD)~\cite{specDrivenDevelopment}. Stil, they lack a structured approach to requirements specification and testing that are crucial factors for ensuring code quality as highlighted in several studies on the integration of LLMs with TDD workflows~\cite{mathews2024test, fakhoury2024llm, piya2025more, ridnik2024code}.

This prior studies validated these TDD workflows leaving the \textbf{human factor} underexplored.
In fact, this presents a \textbf{gap} in understanding the user's role in guiding the LLM through structured requirements specification and test case definition, and, more importantly, how effectively they can express their intent through the overalll TDD-based workflow.
%While many LLM code generation tools rely on chat interfaces, our hypothesis is that a higher level of abstraction can be achieved through a structured TDD workflow allowing users to focus exclusively on requirements.

In this paper, we propose a an empirical experiment to investigate how developers can effectively interact with such a workflow, focusing more on how they can express their intent through the specification and test case definition phases.
% \color{black}
To this end, we rely on \ourTool, a plugin that implements a typical TDD-code generation workflow based on the existing literature~\cite{mathews2024test, fakhoury2024llm, ridnik2024code}, within a popular IDE, Visual Studio Code (VSCode).
\ourTool mainly consists of a three-phase TDD workflow: the \emph{problem description} phase, (1) where the user inputs the specification, (2) the \emph{test cases} phase, where the user guides the LLM in generating and refining a test suite that formally describes the requirements. (3) the final phase, \ie \emph{code generation}, is delegated entirely to the LLM, which generates the code and verifies its execution results against the test suite.
We plan a controlled experiment with human participants, asked to solve a set of programming problems from the LiveCodeBench dataset.
Working exclusively within \ourTool and VS Code, participants will be asked to produce a valid solution for the proposed problem. We will collect detailed data on effectiveness (\eg \emph{PassAll}, \emph{PassRate}, \rev{R1.5}{\emph{TestCoverage}, \emph{TestDiversity}}), efficiency (\eg \emph{TimeToPass}), and interactions (\eg \emph{TestEdits}, \emph{SuiteRegenerations}) in order to analyze the real impact of user involvement in the test curation phase on the overall code generation success.

The expected outcome of our study is twofold: first, to provide empirical evidence on the effectiveness of a TDD-based workflow for LLM-assisted code generation focusing on the user perspective and second, to inform the design of future IDEs, clarifying the trade-off between investing in requirements engineering and code refinement, and how the user can best contribute to the process.

\rev{R3.2}{The rest of the paper is structured as follows. Section~\ref{sec:back} provides background information and related works. Sections~\ref{sec:experimentDesign} and \ref{sec:execution} presents \ourTool and the proposed experiment, while Section~\ref{sec:theatstovalidity} discusses threats to validity and Section~\ref{sec:contributions} summarizes the conclusions.}
%%%%%%%%%%%%%%%%%%%%%%%%%%%%%%%%%%%%
%%%%%%%%%%%%%%%%%%%%%%%%%%%%%%%%%%%%
\section{Background and Related Work}
\label{sec:back}
%%%%%%%%%%%%%%%%%%%%%%%%%%%%%%%%%%%%
%%%%%%%%%%%%%%%%%%%%%%%%%%%%%%%%%%%%

% Several specialized LLMs for code generation have been developed, both closed and open weight models. Some notable closed models are OpenAI's GPT-4~\cite{openai2024gpt4technicalreport} and Anthropic's Claude~\cite{Anthropic2024Claude3ModelCard}. Open-weight models include CodeLlama~\cite{roziere2024codellamaopenfoundation}, StarCoder~\cite{lozhkov2024starcoder2stackv2}, DeepSeek-Coder~\cite{guo2024deepseekcoderlargelanguagemodel}, and the recent Qwen3-Coder~\cite{yang2025qwen3technicalreport}.

% Several benchmarks have been proposed to evaluate the performance of these models by solving coding problems from competitive programming platforms.

A significant contribution in this area was made by Chen \etal~\cite{chen2021evaluating}, who introduced the \emph{Codex} model and an evaluation protocol defining the \texttt{pass@k} metric, the current standard, and the HumanEval benchmark, consisting of coding problems with associated unit tests.

% Additional studies have built upon this foundation, such as SWE-bench~\cite{jimenez2024swebenchlanguagemodelsresolve}, which assesses LLMs' ability to resolve real-world GitHub issues, providing insights into their practical utility in software development. A recent notable contribution is LiveCodeBench~\cite{jain2024livecodebench}, designed to evaluate LLMs on coding problems sourced from competitive programming platforms, implementing a rolling update mechanism to ensure the benchmark remains up-to-date with new problems. This approach mitigates the issue of data contamination, where models might have been trained on benchmark problems themselves.

Other research has examined how LLMs can be combined with established software engineering methodologies to improve code quality and reliability, such as TDD. Mathews \etal~\cite{mathews2024test} proposed the \texttt{TGen} approach, which implements a TDD-guided code generation flow using LLMs. Starting from a set of initial test cases, \texttt{TGen} generates code and iteratively updates it until all tests pass. Their experiments showed that integrating TDD workflows enhances both the accuracy and dependability of code produced by LLMs. Along the same lines, Fakhoury \etal~\cite{fakhoury2024llm} introduced \texttt{TICODER}, incorporating human feedback into the TDD-based code generation loop. 
%\texttt{TICODER} automatically generates tests and code candidates from user requirements, enabling iterative refinement through user feedback. 
Additionally, Piya \etal~\cite{piya2024llm4tdd, piya2025more} explored how combining TDD practices with assistants like ChatGPT improves problem-solving on \emph{LeetCode}-style tasks.

In parallel, recent work has investigated the integration of LLMs within development environments and the dynamics of human–AI interaction. Amershi \etal~\cite{Amershi2019} established fundamental principles for effective human–AI collaboration, emphasizing timely feedback, transparency, and user control—principles embodied in our structured \textit{Specification–Tests–Function} workflow. Sergeyuk \etal~\cite{Sergeyuk2024} and Nghiem \etal~\cite{Nghiem2024} highlight the growing importance of IDE-embedded assistants and the need for transparent, user-centric design in coding tools. 
Complementarily, Crupi \etal~\cite{Crupi2025} and Evtikhiev \etal~\cite{Evtikhiev2023} discuss the limitations of existing automatic evaluation metrics, such as BLEU or \texttt{pass@k}, arguing for interactive and context-aware evaluation frameworks. 
Finally, Gao \etal~\cite{gao2025} point to broader challenges in the use of LLMs for software engineering, including trust, reproducibility, and transparency—issues that our work directly addresses through controlled, in-IDE experimentation.

We build upon these prior works by exploring how a structured TDD workflow, starting from user-defined specifications, can be effectively leveraged for generating correct code with LLMs. 
\rev{R1.1}{Our study focuses on the interaction between the user and the TDD approach itself, an aspect that has been limitedly investigated, with the aim of examining the role of user involvement in test case definition and refinement within TDD.}
\newcommand{\rqOne}{\emph{To what extent is \ourTool able to generate code based on the specification provided by the user?}\xspace}
\newcommand{\rqTwo}{\emph{To what extent does \ourTool allow the user to express their intent through a test suite specification?}\xspace}

\newcommand{\areaOne}{\emph{area 1}\xspace}
\newcommand{\areaTwo}{\emph{area 2}\xspace}
\newcommand{\areaThree}{\emph{area 3}\xspace}

\newcommand{\numProblems}{6\xspace}
\newcommand{\numProblemsPerDifficulty}{2\xspace}

%%%%%%%%%%%%%%%%%%%%%%%%%%%%%%%%%%%%
\section{The \ourTool plugin}
%%%%%%%%%%%%%%%%%%%%%%%%%%%%%%%%%%%%

\begin{figure*}[ht]
	\centering
	\includegraphics[width=0.8\textwidth]{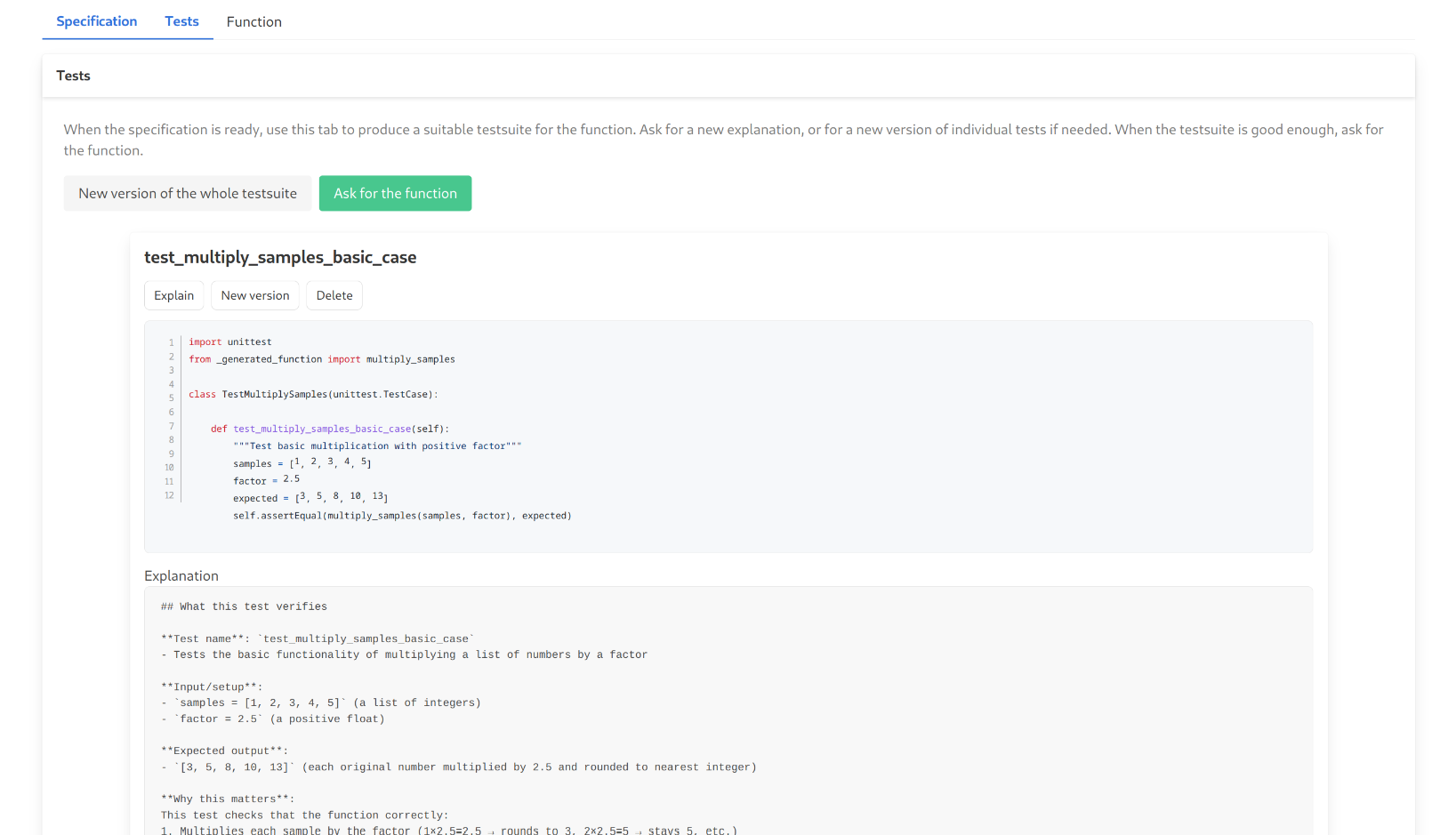}
	\caption{Screenshot of the prototype of \ourTool plugin integrated into Visual Studio Code (VS Code).}
	\label{fig:currante}
\end{figure*}

We designed and developed a Visual Studio Code (VS Code) plugin named \ourTool, which serves as the experimental platform for this study. 
\ourTool implements a TDD workflow that leverages LLMs to generate code based on user-provided specifications formally defined through test cases. 
The workflow is structured into three main phases, represented in the graphical user interface (GUI) as follows:
\begin{enumerate}
    \item \textbf{Problem description (\areaOne):} A text area where the user can input \rev{R1.2}{a structured natural language specification of the problem to be solved. The specification follows a TOML format, capturing the user intent and guiding the LLM to generate the test suite. It also contains the function signature and any necessary constraints or requirements.}
    \item \textbf{Test cases (\areaTwo):} The central section, where the user can review, refine, and improve the test cases describing the problem requirements. An initial test suite is automatically generated from the TOML specification, which serves as input specification for the function generation phase.
    \item \textbf{Code generation (\areaThree):} The final section, which displays the generated code and the corresponding test suite execution results. The user can regenerate the code after refining the test cases.
\end{enumerate} 

The current prototype of \ourTool operates as follows. 
The user starts by defining the problem requirements \rev{R3.2}{in a structured way, via a TOML file. The format is designed to be human-readable and easy to edit\footnote{\url{https://toml.io/en/}}}.
Upon launching \ourTool, the TOML specification is automatically loaded into \areaOne, where it can be further edited. 
Next, the user initiates the test suite generation process, which employs an LLM to produce an initial set of test cases, displayed in \areaTwo. \rev{R1.3}{While the TOML specification provides the initial context for test generation, the test suite itself serves as input specification for the overall Spec-Driven Development workflow.}
An example of the user interface is shown in \figref{fig:currante}. 
The user can iteratively refine the generated test cases by (i) requesting natural language explanations for individual tests, (ii) deleting irrelevant ones, or (iii) providing additional guidance to the LLM to regenerate a single test or the entire suite. 
Once satisfied with the resulting test suite, the user proceeds to the code generation phase (\areaThree), where \ourTool uses the LLM to produce the corresponding code implementation based on the refined tests.

%%%%%%%%%%%%%%%%%%%%%%%%%%%%%%%%%%%%
%%%%%%%%%%%%%%%%%%%%%%%%%%%%%%%%%%%%
\section{Experiment Goal and Research Questions}
\label{sec:experimentDesign}
%%%%%%%%%%%%%%%%%%%%%%%%%%%%%%%%%%%%
%%%%%%%%%%%%%%%%%%%%%%%%%%%%%%%%%%%%

The \textit{goal} of this experiment is to evaluate whether an LLM-based TDD-inspired workflow is effective in generating correct code by providing and refining a formal input specification (i.e., test cases). 
We implemented this workflow in a VS Code plugin, \ourTool, which we use to conduct the experiment. 
The \textit{perspective} is that of software developers and researchers interested in LLM-assisted code generation workflows.

%%%%%%%%%%%%%%%%%%%%%%%%%%%%%%%%%%%%
\subsection{Research Questions}
%%%%%%%%%%%%%%%%%%%%%%%%%%%%%%%%%%%%

The experiment is guided by the following research questions:

\begin{description}
    \item[\RQ{1}:] \rqOne We aim to measure the effectiveness of a TDD-based workflow, implemented through \ourTool, in generating correct code solutions for a given problem without requiring the user to write code directly, but instead by providing and refining a formal specification through test cases. \rev{R1.4}{This RQ focuses more on the validation of the tool's code generation capabilities.}
    \item[\RQ{2}:] \rev{R1.4}{\rqTwo}
    % We aim to evaluate the effectiveness of \ourTool in assisting users in defining a correct and comprehensive set of test cases that capture the problem requirements, thereby enabling the generation of correct code solutions.
    \rev{R1.4}{This RQ focuses more on the human-in-the-loop aspect, measuring how well users are able to express their intent producing an effective test suite that captures the functional problem requirements.}
\end{description}

%%%%%%%%%%%%%%%%%%%%%%%%%%%%%%%%%%%%
%%%%%%%%%%%%%%%%%%%%%%%%%%%%%%%%%%%%
\subsection{Context Selection}
%%%%%%%%%%%%%%%%%%%%%%%%%%%%%%%%%%%%
The context of the study is composed of subjects and objects. 
The subjects are software developers, using \ourTool to solve a set of coding problems.
The objects are coding problems selected from the \emph{LiveCodeBench} dataset~\cite{jain2024livecodebench}.

%%%%%%%%%%%%%%%%%%%%%%%%%%%%%%%%%%%%
\subsubsection{Participants and Settings}
%%%%%%%%%%%%%%%%%%%%%%%%%%%%%%%%%%%%
\rev{R2.3,R3.4}{We aim to recruit participants with a diverse range of experience levels (\ie both junior and senior developers), backgrounds (\ie software developers, computer science students, and professionals from different industries), and demographics.}
All sessions are individual and conducted on a \emph{single-monitor} workstation (desktop or laptop) with keyboard and mouse, using VS Code (fixed version) and the \ourTool extension; the LLM endpoint, prompts, and parameters are fixed across participants. 
Before the tasks, participants receive a brief tutorial and one warm-up example; during the study, they work only within the IDE (no external tools or web search). The extension logs time-stamped actions (produce/explain/regenerate tests; ask/regenerate function; re-run tests) and unit-test outcomes (per-test pass/fail). Participation is voluntary under informed consent; data are pseudonymized and contain only the artifacts and interaction logs necessary for the study.

%%%%%%%%%%%%%%%%%%%%%%%%%%%%%%%%%%%%
\subsubsection{Dataset Selection}
%%%%%%%%%%%%%%%%%%%%%%%%%%%%%%%%%%%%

We plan to select a sample of coding problems from the \emph{LiveCodeBench} dataset~\cite{jain2024livecodebench}, which contains a large set of high-quality coding problems, defined for different coding tasks, extracted from competitive programming platforms. The benchmark is continuously updated with new problems, providing detailed information for each problem, such as test cases, problem description, difficulty level, and solution. 
%We plan to measure the effectiveness of \ourTool across different complexity levels. 
More details are reported in the \emph{HuggingFace} dataset card\footnote{\url{https://huggingface.co/datasets/livecodebench/code_generation_lite}}.
The version of the dataset will be \texttt{release\_v5}, containing a total of 880 coding problems until January 2025. We will select a sample of three short warmup problems (easy difficulty), and three evaluation-candidate problems (medium difficulty). The experiment will be conducted only on the evaluation-candidate problems.
% , ensuring balanced coverage of problem types (\eg algorithms, data structures) and avoiding overlap with known LLM training data (as feasible) considering the reported date for the problems.
% The measurements will be collected only for the evaluation-candidate problem (excluding the warmup problems).
% The selected problems will be validated during the pilot phase of the experiment.

The problem candidates will be selected based on the following criteria:
\begin{itemize}
    \item \textbf{Difficulty level:} We will select problems labeled as \emph{easy} and \emph{medium} difficulty, to ensure a reasonable challenge for participants, while being solvable within the time constraints of the experiment.
    \item \textbf{Diversity of problem types:} \rev{R1.6}{We will select a mix of problem types (\eg algorithms, data structures) to evaluate different coding scenarios, as diverse as possible, with a trade-off ensuring the feasibility of the experiment. A possible approach is to (i) randomly select a problem, (ii) evaluate its type and feasibility, and (iii) keep or discard it until the desired diversity is achieved.} 
    This will apply to the evaluation-candidate problems. The warm-up problems will be selected to be simple and quick to solve, helping to familiarize participants with the tool and workflow.
    \item \textbf{Problem length:} We will select problems that can be reasonably solved within the time constraints of the experiment, \rev{R1.6}{ensuring that the problem description is easily understandable and the required constraints and edge cases are not verly complex}.
    \item \textbf{Avoiding data contamination:} We will avoid problems that are likely to have been included in the training data of the LLM used in \ourTool, based on the reported dates and known datasets.
\end{itemize}

%%%%%%%%%%%%%%%%%%%%%%%%%%%%%%%%%%%%
\subsection{Variables}
%%%%%%%%%%%%%%%%%%%%%%%%%%%%%%%%%%%%

We consider three groups of variables aligned with our \ourTool workflow on \emph{LiveCodeBench} tasks. 
\emph{Independent variables} capture task context (task identifier). 
\emph{Dependent variables} measure effectiveness and process—final correctness (all tests passed and pass rate), efficiency (time to pass), \rev{R1.5}{test suite strength (tests coverage and diversity)} and iteration dynamics (function regenerations, per-test edits, full suite regenerations, and advice triggers). 
\emph{Confounding variables} control for prior skill and habits (years of programming experience, Python familiarity, prior TDD experience, and prior use of LLMs for coding). 
All metrics are computed from the plugin’s time-stamped logs; times are measured from the first produced test suite, and counts aggregate user actions recorded by CURRANTE. 
A complete list with scales is summarized in \tabref{tab:variables}.

\arrayrulecolor{black}
\begin{table*}[ht]
  \footnotesize
  \centering
  \begin{tabular}{p{2.2cm}p{3cm}p{8cm}p{2.4cm}}
    \hline
    \textbf{Variable Type} & \textbf{Name} & \textbf{Description} & \textbf{Scale} \\ \hline

    Independent & TaskId & Identifier of the specific problem instance used in the session (blocking factor). & Categorical (string) \\ \hline

    Dependent & PassAll & Whether the final generated function passes \emph{all} tests in the suite. & Binary: 0 (no), 1 (yes) \\ \hline
    Dependent & PassRate & Fraction of passed tests over total tests for the final submission. & Real number $[0,1]$ \\ \hline
    Dependent & \rev{R1.5}{TestCoverage} & \rev{R1.5}{How much of the source code is executed when the test suite runs.} & Real number $[0,1]$ \\ \hline
    Dependent & \rev{R1.5}{TestDiversity} & \rev{R1.5}{How different the test cases are from one another, using a similarity metric (e.g., Jaccard similarity).} & Real number $[0,1]$ \\ \hline
    Dependent & TimeToPass & Time from producing the first test suite to the first submission that passes all tests (or budget expiration). & Real number (seconds) \\ \hline
    Dependent & IterationsToPass & Number of function regenerations until first all-pass (or budget expiration). & Integer (count) \\ \hline
    Dependent & TestEdits & Number of per-test actions (explain / regenerate / delete) performed by the participant. & Integer (count) \\ \hline
    Dependent & SuiteRegenerations & Number of full test-suite regenerations triggered from the Specification/Tests tab. & Integer (count) \\ \hline
    Dependent & AdviceTriggers & Number of times advice was requested/generated from failing outputs. & Integer (count) \\ \hline

    Confounding & ProgrammingExperienceYears & Participant's general programming experience. & Integer (years) \\ \hline
    Confounding & PythonFamiliarity & Self-reported familiarity with Python. & Categorical: ``none'', ``low'', ``medium'', ``high'' \\ \hline
    Confounding & PriorTDDExperience & Self-reported experience with TDD-like workflows. & Categorical: ``none'', ``low'', ``medium'', ``high'' \\ \hline
    Confounding & PriorLLMCodeGenUse & Prior use of LLMs for coding tasks. & Categorical: ``never'', ``occasionally'', ``frequently'' \\ \hline
  \end{tabular}
  \caption{Variables used in the experiment with \ourTool on LiveCodeBench tasks.}
  \label{tab:variables}
\end{table*}

%%%%%%%%%%%%%%%%%%%%%%%%%%%%%%%%%%%%
%%%%%%%%%%%%%%%%%%%%%%%%%%%%%%%%%%%%
%%%%%%%%%%%%%%%%%%%%%%%%%%%%%%%%%%%%
\section{Experimental Procedure}
%%%%%%%%%%%%%%%%%%%%%%%%%%%%%%%%%%%%
We follow the \emph{ACM SIGSOFT Empirical Standards}~\cite{acm_standard} for quantitative studies involving human participants. 
Before the main task, each participant receives a short interactive tutorial introducing the \ourTool interface and workflow, including the three tabs (\emph{Specification}, \emph{Tests}, and \emph{Function}), and a step-by-step demonstration of how to produce, refine, and validate test suites. 
After this tutorial, participants complete one randomly selected \emph{warm-up task} to familiarize themselves with the process and the experimental environment. 
This training phase ensures that participants understand the workflow and controls for learning effects during the main evaluation.

The experiment \rev{R3.4}{follows a \emph{between-subjects} design}: each participant solves one \emph{LiveCodeBench} problem using the same \ourTool workflow (\emph{Specification} $\rightarrow$ \emph{Tests} $\rightarrow$ \emph{Function}). 
Each participant is randomly assigned one \emph{medium}-difficulty problem from a set of three candidates, ensuring balanced assignment across participants. 
We use the specific \emph{TaskId} as a blocking factor to account for variation among problems.
%\rev{R1.7}{Note that we do not include a baseline comparison in this study, as our focus is on understanding the human-in-the-loop dynamics within the \ourTool workflow, rather than validating its effectiveness against other methods.}

Measurements are collected only during the main evaluation task (excluding the warm-up), capturing user actions, outcomes, and performance metrics. 
This procedure supports a fair assessment of effectiveness and efficiency under a consistent spec-driven process, while satisfying core empirical standards regarding design clarity, control of nuisance variables, and reproducibility.

\rev{R2.1}{The underlying LLM model will be selected based on the most recent and capable open-weight model available at the time of experiment execution. A possible option is \emph{Qwen3-Coder}~\cite{yang2025qwen3technicalreport}, which is specialized for coding tasks. 
The prompts will be kept simple to avoid additional influencing factors (\ie no advanced prompting techniques), following the TDD workflow outlined above.}

%%%%%%%%%%%%%%%%%%%%%%%%%%%%%%%%%%%%
\subsection{Data Collection}
%%%%%%%%%%%%%%%%%%%%%%%%%%%%%%%%%%%%
We instrument \ourTool to capture time-stamped interaction logs and outcome data aligned with the variables in Table~\ref{tab:variables}. 
\emph{Demographics} (pre-study) data collection include years of programming experience, Python familiarity, prior TDD experience, and prior LLM usage for coding; \emph{post-task feedback} captures perceived usability and workload (short Likert items and free-text). 
\emph{Process logs} record every action with timestamps and payload hashes: produce test suite, per-test actions (explain/regenerate/delete), full-suite regenerations, ask/regenerate function, re-run tests, and function viewing events (e.g., when the participant inspects or edits the generated code). 
\emph{Execution results} store per-test pass/fail, failure messages, aggregate pass rate, and whether all tests passed; time-to-pass is computed from the first produced test suite to the first all-pass submission (or budget exhaustion). 
\emph{LLM usage metrics} include token counts for each API call and total tokens consumed per participant, enabling efficiency and cost analyses. 
\emph{Artifacts} (specification in TOML format, test versions, function versions) are stored as text with content hashes to ensure integrity and reproducibility without retaining identifiable project code. 
All logs are pseudonymized, collected on the study workstation, and exported to a CSV/JSON bundle alongside artifact snapshots for analysis.
\rev{R1.9}{Even if participants do not complete the task or do not produce usable tests, we still analyze their usage logs to have insights on potential issues in the \ourTool workflow.}

%%%%%%%%%%%%%%%%%%%%%%%%%%%%%%%%%%%%
\subsection{Coding Tasks}
%%%%%%%%%%%%%%%%%%%%%%%%%%%%%%%%%%%%
Participants will complete the coding task (\ie problem) entirely within VS Code using \ourTool. The in-experiment workflow is:

\begin{enumerate}[(a)]
  \item \textbf{Open specification.} Ensure the TOML specification of the assigned problem is open in the editor (Specification tab).
  \item \textbf{Produce initial tests.} Invoke \textit{Produce test suite} to generate a runnable unit-test file.
  \item \textbf{Refine tests (human-in-the-loop).} Inspect and curate the suite using per-test actions: \emph{Explain}, \emph{Regenerate}, \emph{Delete}; optionally regenerate the whole suite. Aim for a clear, sufficient specification-by-tests.
  \item \textbf{Generate function \& execute.} Use \textit{Ask for the function}; \ourTool \rev{R1.8}{first generates the code function, then} runs the tests and reports per-test results and aggregate pass rate, plus auto-generated advice from failing outputs.
  \item \textbf{Iterate to completion.} If failures remain, either (i) refine the tests or (ii) \emph{Re-generate function} guided by the advice; re-run until completion criteria are met.
\end{enumerate}

\noindent\textit{Completion criteria.} The task finishes when the function \emph{passes all tests} or when the fixed time budget elapses. \textit{Measurement.} Time-to-pass is measured from the first test-suite production; all actions (per-test edits, suite regenerations, function regenerations) and outcomes (per-test pass/fail) are logged.

\noindent\textit{Note.} The concrete wording of the assigned medium–difficulty problem and any minor UI micro-steps will be finalized at experiment execution time, preserving the workflow described above.

%%%%%%%%%%%%%%%%%%%%%%%%%%%%%%%%%%%%
\subsection{Metrics}
%%%%%%%%%%%%%%%%%%%%%%%%%%%%%%%%%%%%
We derive all metrics from \ourTool logs and test executions, aligned with Table~\ref{tab:variables}:

\begin{itemize}[leftmargin=*]
  \item \textbf{Correctness.} \emph{PassAll} (binary: all tests passed at task end) and \emph{PassRate} (final $\#$passed / $\#$total tests).
  \item \textbf{Efficiency.} \emph{TimeToPass} (seconds from first test-suite production to first all-pass submission, or to time-budget expiration), and \emph{IterationsToPass} (number of function regenerations until first all-pass, capped at budget).
  \item \textbf{Process activity.} \emph{TestEdits} (count of per-test actions: explain/regenerate/delete), \emph{SuiteRegenerations} (count of full-suite regenerations), and \emph{AdviceTriggers} (count of advice requests generated from failing outputs).
  \item \textbf{Per-test outcomes.} For each test, pass/fail label and last failure message (used only for advice generation and qualitative inspection).
\end{itemize}

\noindent\textit{Computation and handling.} All timestamps are recorded using the system clock; rapid duplicate clicks are filtered out during logging. If the participant does not achieve an all-pass result, \emph{PassAll} is set to 0 and \emph{TimeToPass} is assigned the full time budget.
Any outliers due to interruptions are flagged and excluded in sensitivity checks. 
We plan to report common descriptive statistics (median and interquartile range for times and counts; proportions for \emph{PassAll}), and use Spearman’s~$\rho$ for exploratory correlation analysis.

\noindent\textit{Note.} The exact operationalization of all metrics (\eg timer start/stop points, aggregation/capping rules) will be finalized at experiment execution time and documented in the study protocol.

\section{Execution Plan}
\label{sec:execution}

The study will proceed in three main phases to ensure methodological consistency and transparency of results: \emph{Preparation}, \emph{Execution}, and \emph{Analysis}. Each phase is described below.

\textbf{Preparation Phase.}
We will freeze the experimental environment by fixing the VS Code version, \ourTool build, prompt templates, and LLM endpoint parameters (temperature, model, and seed) to maximize consistency across runs. The \emph{medium}-difficulty subset of LiveCodeBench will be filtered to remove duplicates and problems potentially included in LLM pretraining corpora. We will select three \emph{training} tasks (easy difficulty) and three \emph{evaluation} tasks (medium difficulty), ensuring diversity in algorithmic type and data structure usage. A pilot session (2--3 participants) will validate the selected problems, timing, LLM configurations, and data collection pipeline. All experimental materials (consent form, pre/post questionnaires, task sheets) will be finalized after the pilot, documenting any changes and their rationale.

\textbf{Execution Phase.}
Each participant will complete the study individually on a single-monitor workstation with VS Code and \ourTool pre-installed. After giving informed consent, participants will complete a short pre-study questionnaire (demographics, programming experience, Python familiarity, prior TDD and LLM-assisted coding). They will then perform: (i) a brief \emph{training task} randomly assigned from the three easy problems to become familiar with the interface; and (ii) one \emph{evaluation task} randomly assigned from the three medium-difficulty problems, using balanced assignment across participants. 

During the evaluation, participants follow a fixed workflow within \ourTool (Specification $\rightarrow$ Tests $\rightarrow$ Function) under a fixed time budget (approximately 30--45 minutes in total). All actions are logged automatically by the extension, including time-stamped interactions, per-test outcomes, token counts, and code inspection events (e.g., when the participant views or edits the generated function). External tools or web search are not permitted. After completing the task, participants will fill out a brief post-study questionnaire to report perceived usability and workload (Likert items and free-text comments). All logs and artifacts (TOML specification, test versions, function versions) are stored as text with content hashes to preserve integrity and anonymity.

\textbf{Analysis Phase.}
We will start with descriptive summaries of all variables listed in Table~\ref{tab:variables} (e.g., proportions for categorical outcomes, medians/IQR for continuous variables, distributions of process actions). 
Further analyses will be chosen \emph{as warranted by the collected data}. Depending on the data characteristics, we may employ non-parametric comparisons for times and counts, basic regression analyses for associations with participant factors, or survival-style summaries if time-to-pass shows right-censoring. 
Any effect-size estimation, multiple-comparison handling, or robustness checks will be applied pragmatically and reported transparently in the final analysis plan. All de-identified logs, scripts, and artifact snapshots will be archived for replication and secondary analysis.

This structured plan ensures a coherent progression from preparation to execution and analysis, minimizing bias while maintaining flexibility to adapt the statistical approach to the actual data collected.

%%%%%%%%%%%%%%%%%%%%%%%%%%%%%%%%%%%%%
%%%%%%%%%%%%%%%%%%%%%%%%%%%%%%%%%%%%%
\section{Threats to validity}
\label{sec:theatstovalidity}
%%%%%%%%%%%%%%%%%%%%%%%%%%%%%%%%%%%%%
%%%%%%%%%%%%%%%%%%%%%%%%%%%%%%%%%%%%%
Threats to validity are potential biases or uncontrolled influences that can distort findings. We distinguish \emph{internal validity} (factors inside the study that affect causal interpretation, e.g., participant heterogeneity, instrumentation, learning effects) from \emph{external validity} (limits to generalizing results beyond our sample, tasks, tools, and setting).

%%%%%%%%%%%%%%%%%%%%%%%%%%%%%%%%%%%%%
\subsection{Internal Validity}
%%%%%%%%%%%%%%%%%%%%%%%%%%%%%%%%%%%%%
\textbf{Subjects.} Participant heterogeneity (programming experience, Python familiarity, prior TDD/LLM use) may influence outcomes. We mitigate this by collecting demographics pre-study and analyzing with these factors as covariates; instructions and time budget are kept constant across sessions.

\noindent\textbf{Task sampling.} Each participant solves one \emph{medium}-difficulty LiveCodeBench task, randomly selected from a set of three possible candidates, treating \textit{TaskId} as a blocking factor. 

\noindent\textbf{Learning and fatigue.} A short warm-up precedes the evaluation task to reduce initial confusion; only one evaluation task per participant limits fatigue and carryover effects.

\noindent\textbf{Instrumentation.} Logging or timing errors could bias metrics (e.g., TimeToPass). We validate the logger in a pilot, store artifact hashes, and compute times from first test-suite production to passing (or budget).

\noindent\textbf{LLM non-determinism.} LLM outputs can vary due to inherent non-determinism.
We will fix the prompt templates and key parameters (e.g., temperature, model endpoint) to maximize consistency across runs. 
Although exact reproducibility cannot be guaranteed in human–LLM interaction studies (due to model stochasticity and user behavior) we expect the outcomes to remain aligned across replications. 
Running several independent sessions allows to average out the variation, similar to individual differences observed in human-subject experiments.

\noindent\textbf{Researcher/experimenter effects.} Standardized instructions and scripted briefings are used; no assistance is given during tasks beyond clarifying the UI workflow.

%%%%%%%%%%%%%%%%%%%%%%%%%%%%%%%%%%%%%
\subsection{External Validity}
%%%%%%%%%%%%%%%%%%%%%%%%%%%%%%%%%%%%%
\textbf{Population.} The sample may not represent all developers. We report demographics and caution when generalizing beyond similar profiles.

\noindent\textbf{Setting and tools.} Results reflect a single-monitor VS Code workflow, Python/unittest, and a fixed LLM. Generalization to other IDEs, languages, or models may be limited. However, the structured TDD-based approach should generalize conceptually, and the model choice will be documented for context.

\noindent\textbf{Task domain.} Using only \emph{medium}-difficulty LiveCodeBench problems constrains scope. The findings may not be generalizable for easy/hard problems, real use-case scenario, such as in industry. 
However, LiveCodeBench provides a diverse set of algorithmic and data-structure tasks that are representative of common coding challenges.
%\rev{R1.6}{Also, the selected representative subset is to optimize the trade-off between diversity and the total experimental duration.}

\noindent\textbf{Ecological validity.} Disallowing external web search or auxiliary tools increases internal control but may under-approximate real workflows; we note this trade-off when interpreting efficiency results.

\noindent\textbf{Sample size.} A modest $N$ limits precision; we prioritize descriptive summaries and only apply simple inferential analyses as warranted by data characteristics.

\noindent\textit{Note.} Additional threats may emerge during the full execution; any newly identified risks and mitigations will be documented in the final study protocol.

\section{Contributions and Implications}
\label{sec:contributions}

This registered report introduces \ourTool, a VS Code extension operationalizing an LLM-assisted, test-first workflow (Specification~$\rightarrow$~Tests~$\rightarrow$~Function) with per-test controls and automatic advice. We contribute: (i) a reproducible task protocol on LiveCodeBench; (ii) a fine-grained telemetry schema capturing time-stamped actions and process metrics; and (iii) a transparent study plan for human-in-the-loop TDD. All materials will be open-sourced for reuse and replication.

For research, this enables principled comparisons of LLM workflows and test curation strategies. For practice, results inform IDE design for safe \emph{spec-by-tests} development. For education, \ourTool provides scaffolded TDD to improve feedback cycles. Broadly, our methodology offers a reusable blueprint for evaluating emerging LLM coding tools with a focus on human–AI interaction and realistic constraints.

\section*{Acknowledgment}
The study presented in this paper was funded in part by the ADVISE project, funded by the Spanish AEI with reference 2024/00416/002.

\bibliographystyle{IEEEtran}
\bibliography{paper}

\end{document}